**Anticorrosion properties of silica-based sol-gel coatings on steel – the influence of hydrolysis and condensation conditions**

Jolanta Gąsiorek[1*], Anna Gąsiorek[1], Bartosz Babiarczuk[1], Walis Jones[2], Wojciech Simka[3], Jerzy Detyna[1], Jerzy Kaleta[1], Justyna Krzak[1]

[1] Faculty of Mechanical Engineering, Wroclaw University of Science and Technology, 54-370 Wroclaw, Poland

[2] BioPharm Enterprises Limited, 65 Gwaun Afan, Cwmafan, Port Talbot, United Kingdom

[3] Faculty of Chemistry, Silesian University of Technology, 44-100 Gliwice, Poland

[*]Corresponding author: jolanta.gasiorek@pwr.edu.pl, justyna.krzak@pwr.edu.pl

**Abstract:** In this paper, we present indirect methods to define the change in the rates of hydrolysis and condensation reactions of a silica oxide network formed of (3-glycidoxypropyl)methyltriethoxysilane (GPTMS) and (3-aminopropyl)triethoxysilane (ApTEOS), which was strongly dependent upon the nature of the solvent used in the reaction environment, such as methanol, ethanol, propanol, butanol and acetone. Electron microscopy (SEM), focused ion beam scanning electron microscopy (FIB), and Raman spectroscopy were used to investigate the morphology and structure changes. UV-VIS spectrometry, a wettability test, a scratch test and electrochemical tests, namely - Tafel's polarization and electrochemical impedance spectroscopy (EIS), were used to analyze the functional changes in the obtained silica coatings. The coating synthesized in a methanol environment showed a polarization resistance ($R_p$) almost an order of magnitude greater than those synthesized in the other organic solvents. The greatest thickness, nearly 1.5 μm, was obtained with the ethanol-based coatings. Almost twice as much force (near 11 N) was required to detach the acetone-based coatings, compared to the other coatings. Propanol and butanol contribute to the defects created and the decreased protection properties in the final materials and show a statistically significant influence on the formed corrosion film under sol-gel coatings, particularly for coatings based on propanol. No statistically significant influence of the solvents was observed on the thickness of coatings prepared with butanol or acetone.

## 1. Introduction

Coating materials protection is the largest sector associated with anti-corrosion prevention. It is estimated that 89% of the costs of anti-corrosion prevention are attributable to the coatings market. [1]. According to the estimated cost of corrosion is about 3.4% GDP [2]. At the same time, the successful anti-corrosion materials must be environmentally friendly and meet environmental standards. One of the most promising methods for using environmentally friendly anticorrosion coating materials is the sol-gel method. Using this green ecological synthesis approach, it is possible to obtain non-toxic and chemically-inert sol-gel materials. Moreover, the sol-gel method used to obtain the coatings does not generate waste products that must either be removed or treated. Synthesis is usually prepared at room temperature, without the input of additional energy [3–6]. Sol-gel materials are becoming increasingly popular across a wide range of potential applications, such as in optical [4], automotive [7], aircraft [8], medicine [9], chemical industries [10], and other application areas [5].

-In order to derive sol-gel materials with specific chemico-physical properties, the influence of the synthesis parameters on the structure of the final product is of key importance. Not only does the nature of the chemical precursors themselves determine the properties of the coatings [5,6,11], but also the types of catalyst and solvent used for synthesis [7,12], as well as reaction temperature [10], aging time of the sols [7], stabilization temperature [13,14], and finally, the number of layers deposited [6,12]. The composition of the sol is the most important factor, since this drives the hydrolysis and condensation reactions and, in this way is ultimately the key influencer of the structural properties of the product material [14].

The type of catalyst determines the growth and size of sol-gel particles [15–18]. Other research points to the fact that the catalyst influences also the thickness of sol-gel coatings [19]. The type of catalyst used, its strength, dissociation degree, the possibility of creating a complex, make it possible to control hydrolysis and condensation reactions [7,10,20]. Decreasing the pH of the sol-gel reaction medium near to the isoelectric point of $SiO_2$ results in a decrease in the kinetics of the condensation reaction, giving rise to more dense films through controlling pore size distribution [14,20]. Additionally, some catalysts, such as acetic acid, can result in increased stability of the sols, for example, by creating chemical complexes, by preventing the sols from precipitating as particles. [7]. It is also possible that a high concentration of catalyst is able to promote the corrosion of the metallic substrate in itself during the deposition process. This, in turn, can result in the decreased adhesion of the coating to the substrate by interfering and, therefore, limiting the ability of reactive chemical species, such as Si-OH free groups in the sol, to form homogeneous surface bonding to the substrate [7]. In this way, it is possible that such catalyst-induced corrosion can serve as foci for subsequent environmentally-induced corrosion.

On another level, it can be seen that a critical parameter is recognizing the nature of the interaction between the solvent and the solute, such as the intermolecular forces and energy barriers that exist between the reactants, the degree of polarity and energy input that exist between polar and non-polar molecules, and the influence of the dielectric constant on the overall chemical equilibrium, where increasing the dielectric constant causes a decrease in the force between the molecules during the formation of the structure [4,14,21,22]. Because the by-products of the synthesis are water and alcohol, selection of appropriate solvents may significantly influence the end-structure of sol-gel materials. According to the literature [4,14,18], the solvent phase acts as an agent between siloxanes and water, promoting the uniform dispersal of the reactants during the hydrolysis and condensation reactions of the sol-gel precursors and enables a faster rate of water evaporation. Using a solvent with a high dielectric constant restricts intermolecular reactions and, as a result, decreases the kinetics of the condensation, thereby decreasing the degree of conversion of the reactants to the gel form [14]. Moreover, aprotic solvents are more inert to the type of chemical reactions that drive the sol-gel process [4]. Based on their use of a polar and protic solvent (ethanol), Khan et al. [4] obtained a stiffer and condensed sol-gel structure with small pores, while the opposite was observed with an aprotic solvent, resulting in decreased homogeneity and an increase in the pore-size of the matrix. Additionally, their research suggests that aprotic solvents promote irreversible hydrolysis and condensation reactions [4]. According to [5,23], the solvent has an influence on microstructure and film crystallization, the size of crystallite and its properties as a result slow the rate of hydrolysis and the kinetic of condensation. Other research [24] suggests that increasing the polarity of the reaction medium modulates cation distribution in the crystalline network, promoting the growth phase. The limited solubility of the precursors in the solvent phase during the final gelation stage can result in the creation of secondary oxide forms, in the form of corrosion products, such as oxyhydroxides and oxides, on the surface of the iron substrate. Furthermore, more organic solvents such as alcohols, with a hydrocarbons moieties, provide energy during calcination, which can be used to promote the crystallization reaction. The energy from the consumption of hydrocarbons (from the solvent) decreases the crystallization temperature of the film and reduces the grain-size of the crystallites (longer hydrocarbon chains reduce the grains of the crystallite) [22]. Other research shows the influence of ethanol and methanol on the opening of the epoxy ring of GPTMS, which influences the density of the sol-gel film and, as a result, increases the hardness of the hybrid sol-gel films that are formed [10,11].

This paper shows the influence of five different solvents on changes in the properties of silica sol-gel coatings based on GPTMS (3-glicydoxypropyltrimethosysilan) and ApTEOS (3-aminopropyltriethosysilan). Methanol, ethanol, propanol, butanol, and acetone were used as solvents in the sol-gel reactions. Scanning electron microscopy (SEM) and energy dispersive x-ray (EDX,) Raman spectroscopy, and wettability tests were used to characterize the physicochemical changes in sol-gel materials. Potentiodynamic polarization and electrochemical impendency spectroscopy (EIS) were used to define the protection properties of the sol-gel coatings in 0.5 M NaCl. The scratch test was used to characterize changes in mechanical properties of the obtained materials.

## 2. *Preparation and Methods*

P265GH steel squares (20x20x3 mm) served as substrates for this study. The composition of the steel is presented in Table 1. All substrates were polished, ending with a final treatment using 800-grade SiC sandpaper. Substrates were then cleaned ultrasonically in acetone and subsequently treated with 0.15% $HNO_3$ (POCH, Gliwice, Poland) in ethanol (96%, POCH, Gliwice, Poland).

**Tabel 1.** P265GH steel composition (according to the PN/EN 10027-2 standard)

| Element | C | Mn | Si | P | S | Cr | Mo | Ni | Al | Cu | Nb | Ti | V | Fe |
|---|---|---|---|---|---|---|---|---|---|---|---|---|---|---|
| % wt. | 0.20 | 1.40 | 0.40 | 0.025 | 0.02 | 0.30 | 0.08 | 0.30 | 0.02 | 0.30 | 0.01 | 0.04 | 0.02 | 96.885 |

Organically modified silica sol-gel materials were prepared using (3-glycidoxypropyl)trimethoxysilane, (97% purity, GPTMS, Alfa Aesar, Kandel, Germany), and 3-aminopropyltriethoxysilane, (98% purity, ApTEOS, Alfa Aesar, Kandel, Germany) as precursors.

Five different solvents were also used. Four of them were protic solvents (methanol, ethanol, propanol, butanol), while one was aprotic (acetone), due to their specific properties to promote the coating of materials for corrosion mitigation [24]. Ethanol (96%, POCH, Gliwice, Poland) or methanol (POCH, Gliwice, Poland) or 1-propanol (POCH, Gliwice, Poland) or 1-butanol (POCH, Gliwice, Poland) or acetone (POCH, Gliwice, Poland) were used. Table 2 summarizes the chemical properties of the solvents used to synthesis the sol-gel.

**Table 2**. Molecular size, dielectric content, and polarity index of solvents used in sol-gel synthesis [25–27]

| SOLVENT | MOLECULAR SIZE CRITICAL DIAMETER (nm) | DIELECTRIC CONSTANT | POLARITY INDEX (p) |
|---|---|---|---|
| Methanol | 0.42 | 33.0 | 6.6 |
| Ethanol | 0.45 | 25.3 | 5.2 |
| Propanol | 0.46 | 20.1 | 3.4 |
| Butanol | 0.46 | 17.8 | 3.9 |
| Acetone | 0.48 | 20.7 | 5.4 |

Sols were catalyzed with $HNO_3$ (65%, POCH, Gliwice, Poland) and $CH_3COOH$ (99.5%, Warchem, Warszawa, Poland).

the molar ratio of GPTMS:ApTEOS was 1:1, the volume ratio of the precursors:solvent was 1:3, and the volume ratio of the precursors:$HNO_3$:$CH_3COOH$ was 31:3:1. Sols homogenized by mixing on a magnetic stirrer for 2 h at room temperature (21 ± 1°C). Synthesis of the sols is illustrated in Fig. 1. Sols were deposited using a dip-coating method to prepare a coating consisting of 3 layers of individual materials. The following procedure was used:

- dipping rate: 180 mm·$min^{-1}$,

- immersion time: 60 s,

- pulling rate: 34 mm·$min^{-1}$,

- drying after deposition of each of layer: 50°C by10 min in a laboratory dryer, (POL-EOKO-APARATURA SP.J., SL 53 STD, Wodzisław Śląski, Poland).

Subsequently, the samples were stabilized at 120°C·3h using a temperature gradient of 0.5°C·$min^{-1}$ (laboratory dryer, POL-EOKO-APARATURA SP.J., SL 53 STD, Wodzisław Śląski, Poland).

In this paper, the following abbreviations and terms have been adopted:

- *'MeOH'* – for coatings based on methanol as solvent,
- *'EtOH'* – for coatings based on ethanol as solvent,
- *'PrOH'* – for coatings based on propanol as solvent,
- *'BuOH'* – for coatings based on butanol as solvent,
- *'Acetone'* – for coatings based on acetone as solvent.
- *Sol-gel* – the method used to prepare the coatings
- *Sol* – liquid phase – the product of the sol-gel method used to coat a substrate
- *Coating* – the product obtained from sol deposition on the substrate and after stabilization

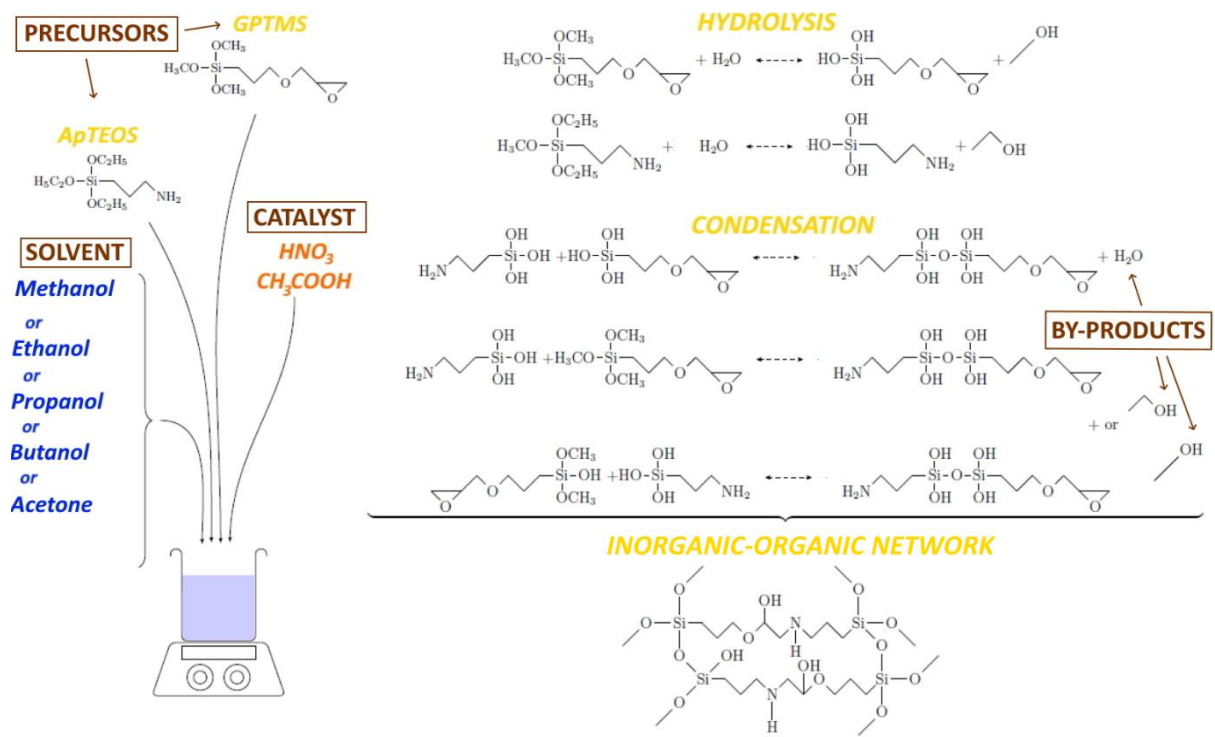

**Figure 1.** Schematic of the synthesis scheme for the sols, illustrating the possible hydrolysis and condensation reactions and the silica network obtained via epoxy ring opening.

The prepared specimens were tested to determine the physicochemical properties of the obtained coatings. For this purpose, a scanning electron microscope (SEM) (S-3400N, Hitachi, Tokyo, Japan) with an energy-dispersive X-ray spectroscopy (EDX) detector (Thermo Scientific, NORAN System Six), a focused ion beam scanning electron microscope (FIB) (SEM/Xe-PFIB, FEI, Massachusetts, USA), a Raman spectrophotometer (LabRAM HR800 HORIBA JOBIN YVON, Lonqjumeau, France), a force tensiometer (OEG GmbH, Frankfurt, Germany) and a UV-VIS spectrophotometer (ThetaMetrices FR-pRO/UV-VIS) were used. The SEM and EDX analyses of the samples were performed using secondary electron (SE) and backscattered electron (BSE) detectors.

The Raman spectra were measured in the range of 100-4000 cm$^{-1}$ using a source of raised oscillation, using an argon laser of 514.5 nm wavelength. The most representative spectra were analysed and presented. Six repetitions of the measurements were conducted for each sample.

The UV-VIS spectra were carried out in the range of 200-800 nm. The measurements were performed using a Thetametrisis FR-pRo / UV-VIS spectrophotometer with halogen and deuterium lamps. Measurements were performed in reflectance mode using the White Light

Reflectance Spectroscopy (WLRS) methodology [28]. At least 14 measurements were conducted for each sample.

The Owens-Wandt-Rabel-Kaelble (OWRK) method was used to establish the polar and dispersive components of the surface free energy (SFE) of the sol-gel coatings. In the measurements, water and mono ethylene glycol were used. For water the (polar part is 51.00 mN·m$^{-1}$ and nonpolar part is 21.80 mN·m$^{-1}$, for mono ethylene glycol polar part is 19.00 mN·m$^{-1}$ and non-polar part is 29.00 mN·m$^{-1}$). All chemicals were analytical grade. For each sample, 6 repetitive measurements were conducted, using a drop volume of 0.1 μl.

The progressive scratch test was carried out using a Micro Combi Tester (MCT), (CSM Instruments, Graz, Austria). A Rockwell tip with 100 μm diameter was used to form 3 mm scratches on each coating (3 repetitions for each sample). A load in the range of 30 mN to 20 N was used, with a linearly increasing speed of 1.5 mm/min. The test was prepared according to the PN-EN 1071-3:2007 standard, where the normal force ($F_n$), friction coefficient (μ), penetration depth ($P_d$) (the depth of the scratch during the increase in a load) and residual depth ($R_d$) (the permanent depth of the scratch track after the load release and elastic recovery of the coating) were determined. Based on $F_n$, μ and $P_d$, the characteristic loads ($L_C$) were determined ($L_{C1}$ - cracking of the coating, $L_{C2}$ - characteristic chipping of the coatings, $L_{C3}$ - penetration of the coating into the substrate in the middle of the scratch).

The electrochemical properties of the materials were characterized with a three-electrode system using a platinum electrode as a counter electrode, a saturated calomel electrode (SCE) as a reference electrode, and the samples as the working electrodes. Potentiodynamic testing (Tafel polarization) and electrochemical impedance spectroscopy (EIS) were performed in a 0.5 M NaCl solution at room temperature (25°C) (ATLAS 1131&EU, AtlasSollich, Rębiechowo, Poland). The open-circuit potential (OCP) was recorded over a period of 2 h, until changes were lower than 5 mV by 10 min. The potentiodynamic polarization was measured in the potential range of -300 mV to +300 mV, with respect to the OCP versus SCE, with a scan rate of 1 mV·s$^{-1}$. Polarization resistance (Rp) was calculated based on the corrosion current densities ($j_{corr}$) and the cathodic and anodic Tafel's slopes ($B_c/B_a$) by using equation (1).

$$R_p = \frac{Ba \times Bc}{(|Ba|+|Bc|)} \times \frac{1}{J\,corr} = \frac{B}{J\,corr} \quad (1)$$

The EIS measurement was conducted over a frequency range from 10 mHz to 100 kHz with an amplitude of 5 mV and 8 points/dec. AtlasLab software (AtlasSollich, Rębiechowo, Poland) was used to fit the EIS data. For each set of samples, 3 repetitions were performed. The measured area was 0.75 cm$^2$.

## 3. Results

The surface morphology of P265GH steel and sol-gel coatings based on various solvents was examined using scanning electron microscopy. The SEM images of steel and coatings surfaces are presented in Fig. 2.

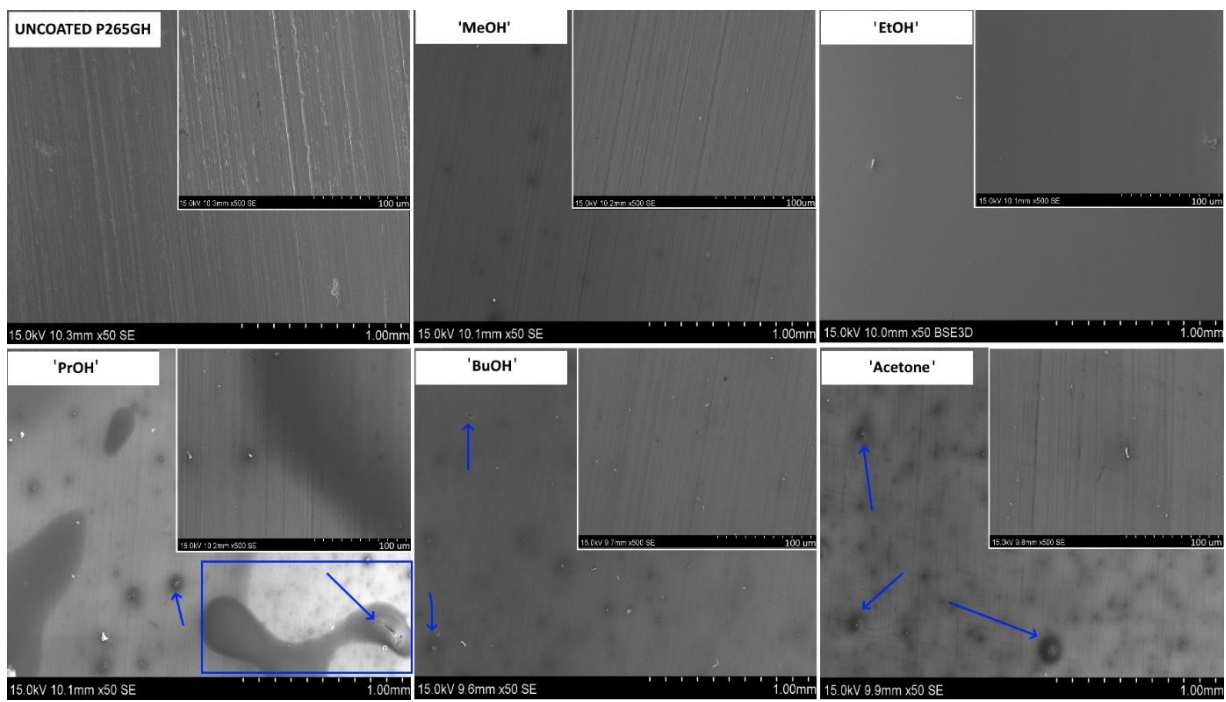

**Figure 2.** SEM images of uncoated P254GH steel and sol-gel coatings based on various solvents

The P265GH steel surface was characterized by scratches, which was the result of pretreatment (results after polishing). The nature of the surface coatings based on the use of the methanol solvent appeared smooth and without cracks or damage. Coatings based on ethanol as the solvent were thicker and denser than the other samples presented and the scratches typically formed on the steel substrate were not observed, probably reflecting the influence of the greater thickness of this coating. For the propanol-based coating, cracks were observed, along with flakes/precipitations. In addition, similar cracks and precipitations were also observed for coatings based on butanol and acetone. The assumptions that ethanol-based coatings were characterized by greater thickness were confirmed by thickness measurements, as presented in Fig. 3. The coatings based on ethanol as a solvent were 1.5 µm, whilst the thicknesses of the other coatings were no more than 1.0 µm. The smallest thickness was observed for the propanol-based coating, at about 630 nm.

The normality of the thickness distribution for each group of sol-gel coatings was checked using the Shapiro-Wilk test [29]. For all solvents, the probability (p-value) in the Shapiro-Wilk test (Table 4) was above the 0.05 significance level, as confirmed by the normal distribution of the obtained results. In the Levene's test [30], the lack of homogeneity of variance (F = 11.56) was found at the probability level, p, equal to $2.9 \cdot 10^{-7}$. Based on the data presented in Fig. 3, it is concluded that the values of the standard errors for individual solvents differ significantly, which confirms the result obtained in the Leven's test. The greatest variation in thickness were obtained for methanol. The reason for such scattering of results may arise from a greater tendency for the sol based on methanol to fill the depressions on the substrates, as a result of which the scratches formed during the polishing pretreatment were covered by a thinner layer of sol-gel coating. Results can be correlated with different viscosities of solvents (methanol 0.5 mPa·s, ethanol 1.1 mPa·s, propanol 2.3 mPa·s, butanol 2.6 mPa·s and acetone 0.3 mPa·s [31,32]) using to sol-gel synthesis, an thanks them various tensions surfaces of sols and as a results various wettability properties.

The Welch F test with variance analysis [33] was used to determine the significance of the influence of solvent on coating thickness. The result of the presented analysis showed a statistically significant influence of the type of solvent on the thickness of the obtained Significant Difference test, it was found that there was no statistically significant difference (for α = 0.05) in thicknesses between coatings based on acetone and butanol. For all other solvents used for synthesis, statistically significant differences were observed between them in the thickness of the coatings (p-value is lower than the assumed significance level of 0.05), as shown in Table 3. The thickness of the sol-gel coatings increases with the increasing viscosity of the solvents: acetone<methanol<ethanol. However, a different trend was observed for propanol and butanol, where the thickness of the coating drops, despite the increased viscosity of the solvents. For the validation of the results the thickness for coatings based on ethanol the focused ion beam microscope (FIB) was used and the results are shown in Fig. 4. In order to the improve imaging quality, the samples were coated with platinum (observed as grey top layer). As can be observed, the system consists of only one layer having a thickness of 0.5 µm and the three-layer system were about 1.5 µm, respectively.

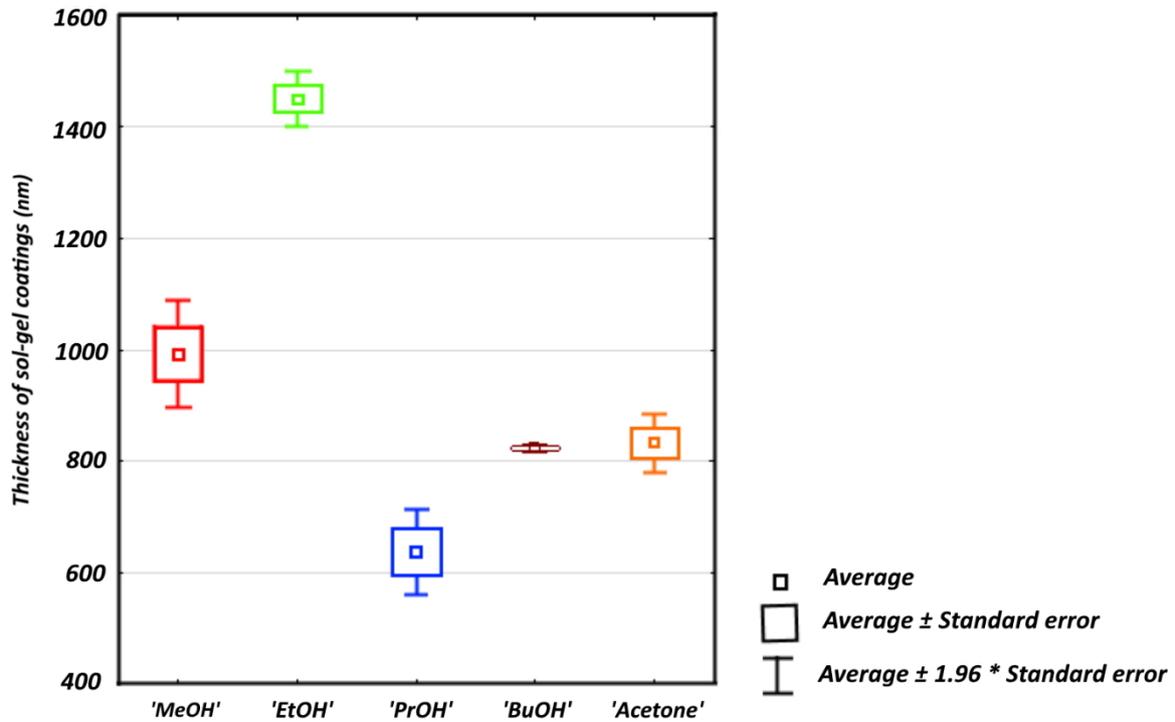

**Figure 3.** Average thickness of sol-gel coatings with marked standard errors and a confidence interval of 95.00%.

**Table 3.** P-values obtained from the HSD Reasonable Difference test of sol-gel coatings

|  | 'MeOH' | 'EtOH' | 'PrOH' | 'BuOH' | *'Acetone'* |
|---|---|---|---|---|---|
| *'MeOH'* |  | 0.000125 | 0.000125 | 0.009264 | 0.016341 |
| *'EtOH'* | 0.000125 |  | 0.000125 | 0.000125 | 0.000125 |
| *'PrOH'* | 0.000125 | 0.000125 |  | 0.003271 | 0.001794 |
| *'BuOH'* | 0.009264 | 0.000125 | 0.003271 |  | **0.999702** |
| *'Acetone'* | 0.016341 | 0.000125 | 0.001794 | **0.999702** |  |

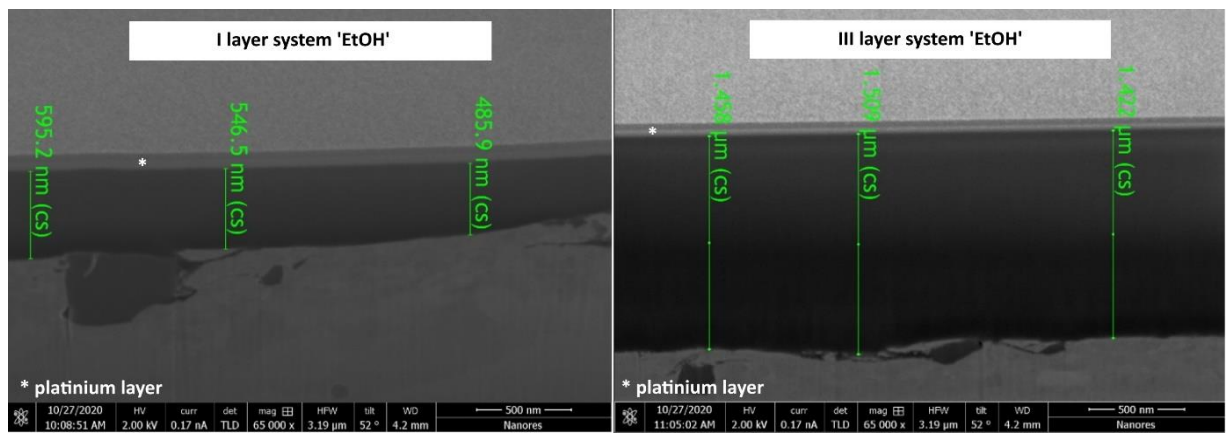

**Figure 4.** FIB micrographs for sol-gel coatings based on ethanol as solvent ('EtOH')

Fig. 5 and 6 shown Raman spectra for the sols which were used to coat the P265GH steel. A band was observed at 209 cm$^{-1}$ for the sols based-on methanol and propanol from several contributions from deformation vibration of propyl groups (from ApTEOS) and the creation the silicooxygen rings (as quadripolar) A band was observed at 475 cm$^{-1}$, indicative of the breathing mode of the oxygen atoms within the silicooxygen rings, was observed for sols based on methanol, ethanol, propanol and butanol [34].

A further band at 581 cm$^{-1}$, arising from the siloxanes ring structure (from moving the silicon in phase) was observed for the sol based on acetone as a solvent [34]. However, the intensity of the bands suggests that ethanol promotes the creation of the ring structure of siloxanes the most efficiently, with the other solvents showing a successively smaller influence in the following order, namely butanol> propanol> methanol. of the changes observed in band intensity at 475 cm$^{-1}$ reflect a more efficient hydrolysis and condensation reaction in ethanol as solvent.

Due to the formation of H$_2$O, methanol and ethanol, also, as a by-products of the hydrolysis and condensation reactions of GPTMS and ApTEOS , it was expected that the use of methanol as a solvent would increase the kinetics of the hydrolysis reaction beyond that of ethanol [14]. Interestingly, despite the expected increase kinetics of the hydrolysis reaction in methanol, the condensation was, in fact, the slowest. This can be explained by the fact that the methanol solvent has a large dielectric constant. Therefore, even if the hydrolysis reaction was to proceed, the reagents cannot easily undergo condensation, due to the nature of the intermolecular reactions [14]. It appears that the kinetics of condensation of sol-gel substrates, observed as the creation of the silicooxygen structure, decrease with the decrease in the length of the protic solvent chains of the solvent used. All of the above data forms a rationale to explain the greatest thickness of coating based on ethanol (≈1.5 µm) – nearly twice the layer thickness observed with the other solvents.

The intensity of the band at 648 cm$^{-1}$ for the sol based on the use of acetone as a solvent is directly related to the amount of unhydrolyzed ApTEOS, indicative of the stretching vibrations of the SiO$_3$ moieties of ApTEOS [35]. However, none of the sols displayed the characteristic doublets at 613 cm$^{-1}$ and 643 cm$^{-1}$ that would have indicated the presence of unreacted or free GPTMS, suggesting that GPTMS was hydrolyzed in each of the sols prepared in this study [35–37].

Additionally, a band at 1256 cm$^{-1}$, indicative of a symmetric ring stretch of the epoxy group from GPTMS, was not observed for all sols, suggesting that the epoxy rings of the GPTMS in these sols were reacted and opened [35–39].

A medium intensity broad band with two maxima at 1041 cm$^{-1}$ and 1107 cm$^{-1}$, respectively, indicative of the aminopropyl segment and the ring form of silicooxygen, were observed for all sols, as shown in Fig. 5 [34].

Broadening signals come from several vibrating modes that overlap each other, namely, from the CH$_2$, -NH, and Si moieties vibrating in the ring [34]. The band at 1193 cm$^{-1}$ arises from the rocking vibration in the C-H links in the Si-CH$_2$ groups [40]. The medium signals observed for all sols at 1296 cm$^{-1}$ and 1456 cm$^{-1}$, respectively, arise from the presence of twisted -CH$_2$ and bending -CH groups from the alkyl chain [41,42].

Together, these observed signals suggest the cross-linking of organic moieties within the precursor mixes involving the opening of the epoxy rings from GPTMS and bonding with amine groups from ApTEOS. Additionally, when the intensity of the bands at 1296 cm$^{-1}$ and 1456 cm$^{-1}$ are compared, it can be seen that the largest band intensities are from the sols based-on methanol and propanol, meaning that hydrolysis and condensation reactions of organic moieties were most effective in these two solvents. at the same time, it can be seen that acetone created the least favorable environment for the hydrolysis and condensation reaction of GPTMS and ApTEOS. The weak band at 1640 cm$^{-1}$ comes from the NH$_2$ moiety [42]. The presence of bands from NH$_2$ could indicate that part of the ApTEOS had not reacted, although signals for unreacted GPTMS were not observed. On the other hand, weak bands at 3316 cm$^{-1}$, indicative of vibration from NH$_2$ groups confirm the polymerization of ApTEOS with GPTMS (Fig. 6), as previously described [42,43]. From the data, it appears that ApTEOS was used in excess in the chemical synthesis, despite the use of a molar ratio to ensure a 1:1 stoichiometry in the original reaction mixture. At 2903 cm$^{-1}$ for sols based on protic solvent (methanol, ethanol, propanol, and butanol) (Fig. 6) observed strong bands assigned to CH$_2$ were observed stretching from the propyl chain [41]. For sols based on aprotic acetone signals with maxima at 2845 cm$^{-1}$ and 2938 cm$^{-1}$ were observed, indicative of CH$_3$ stretching from methoxy moieties present in the matrices [41,44]. These results would suggest that part of the GPTMS within the sol remains unhydrolyzed.

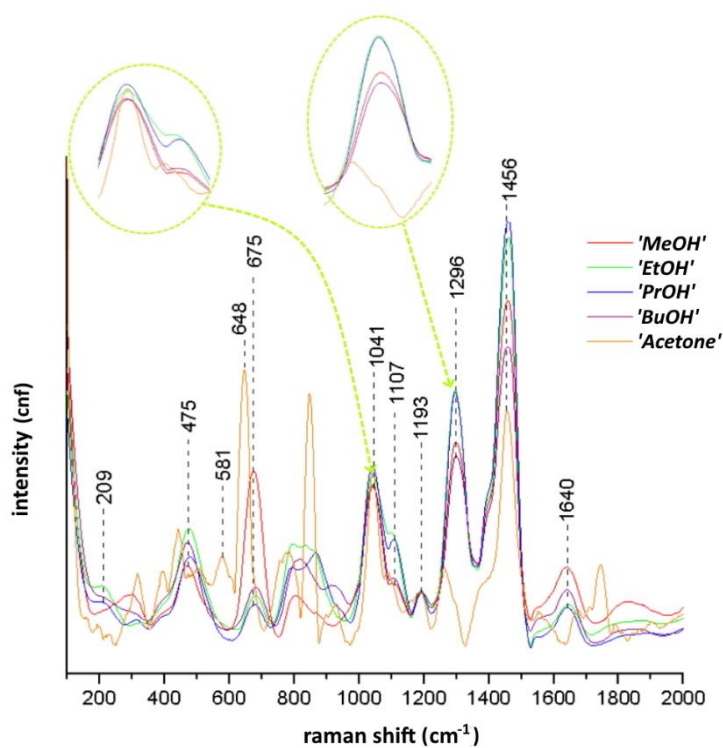

**Figure 5.** Raman spectra of sol-gel sols in the 100-2000 cm$^{-1}$ range

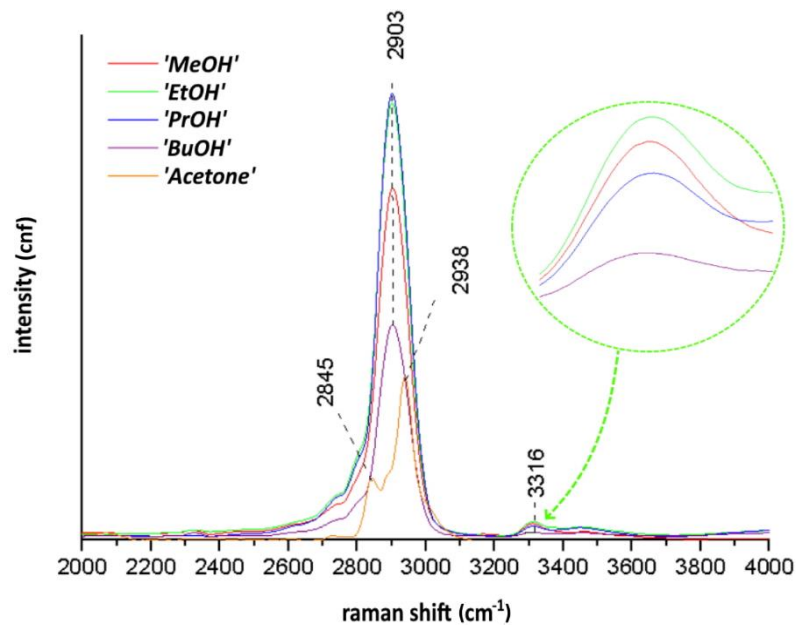

**Figure 6.** Raman spectra of sol-gel sols in the 2000-4000 cm$^{-1}$ range

Fig. 7 presents Raman spectra of coatings based-on various solvents obtained by the sol-gel method. Characteristic bands were observed for the propyl groups from ApTEOS and the creation of quadripolar silicooxygen rings at 209 cm$^{-1}$ [34] for coatings based-on EtOH, and from the aminopropyl segment and the ring-form of silicooxygen at 1041 cm$^{-1}$ and 1107 cm$^{-1}$ [34] for coatings prepared from all solvents. The band observed at 1198 cm$^{-1}$ indicates the rocking vibration between the C-H components within the Si-CH$_2$ groups [40]. Additionally, the medium-sized signals at 1296 cm$^{-1}$ and 1456 cm$^{-1}$ that were observed for all coatings arise from twisting in the CH$_2$ groups and the bending of the CH groups from the alkyl chains [41,42]. Weak bands at 1640 cm$^{-1}$ and 3316 cm$^{-1}$, respectively, come from the NH$_2$ moiety [42]. The more localized additional bands, with weak vibrations, at 492 cm$^{-1}$, 690 cm$^{-1}$ and 1414 cm$^{-1}$, result from Si-O-Si vibrations [36–38], the rocking vibration within the CH$_2$ component from the Si-CH$_2$R groups [45] and the vibrations in the CH$_2$ component from the propyl chains [41], respectively. The bands observed between 2690-3129 cm$^{-1}$, with maxima at 2869 and 2930 cm$^{-1}$, come from the propyl chain of ApTEOS [41]. The profile of the Raman spectral bands confirmed the presence of a silica network modified by an organic network. No bands were observed that would have indicated the presence of the unhydrolyzed precursor (band loss at 648 cm$^{-1}$, 2845 and 2938 cm$^{-1}$ [41]).

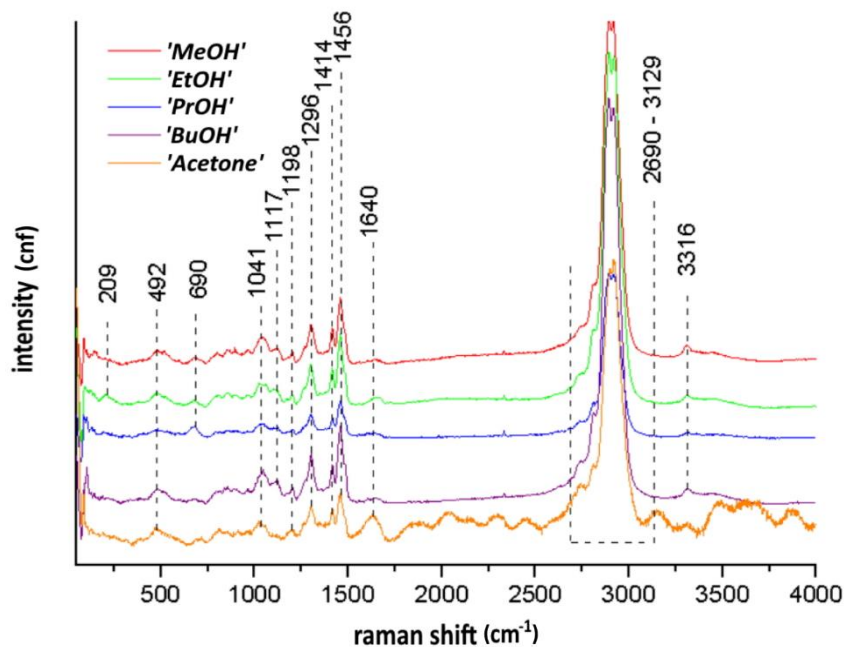

**Figure 7.** Raman spectra for sol-gel coatings based-on various solvents

Fig. 8 and 9 present the contact angle values for water and mono ethylene glycol, along with the surface free energy (SFE) and the polar and dispersive components for uncoated P265GH steel and samples of this steel with sol-gel coatings based on ethanol/methanol/propanol/butanol/acetone solvents. All the surfaces tested were hydrophilic, since the contact angles obtained did not exceed 90°C. However, coated surfaces, where methanol, ethanol and propanol were used as the solvents for synthesis, were characterized by a decreased hydrophilic effect and a corresponding increase in the value of the contact angle for water.

For all coatings, the mono ethylene glycol contact angle was higher than that of the control uncoated steel, indicating that the modification by sol-gel coatings increases the phobic effect on the nonpolar liquid. The results can be explained by the lower share of hydroxyl groups on the surface [46], in this case, the presence of hydrophobizing Si-O-Si moieties [47], which is reflected in the decrease observed for the polar part of SFE (Fig. 9). For all the surfaces analyzed, the dispersive component demonstrated a higher share in the total SFE, but there was less of a share in the Van der Waals interaction on the modified surfaces compared to the uncoated P265GH steel.

The SFF for propanol-based coatings was similar to that of uncoated P265GH steel. However, the ratio of polar to dispersive components in SFE for coatings based on propanol was significantly different. These results can be correlated with the degree of conversion of the sol-gel synthesis substrates [47,48]. On the one hand, the decrease of the polar part in the coatings based on propanol suggests that the materials consist of a low number of Si-OH groups capable of creating hydrogen bonds [47]. On the other hand, the higher ratio of the dispersive component suggests that London forces exert a dominant effect in modifying the properties of the coating materials. For coatings based on methanol and ethanol, the polar, dispersive parts and SFE have similar values, suggesting that the polar solvents, such as methanol and ethanol, have a similar influence on the hydrolysis and condensation reaction between GPTMS and ApTEOS [47,48].

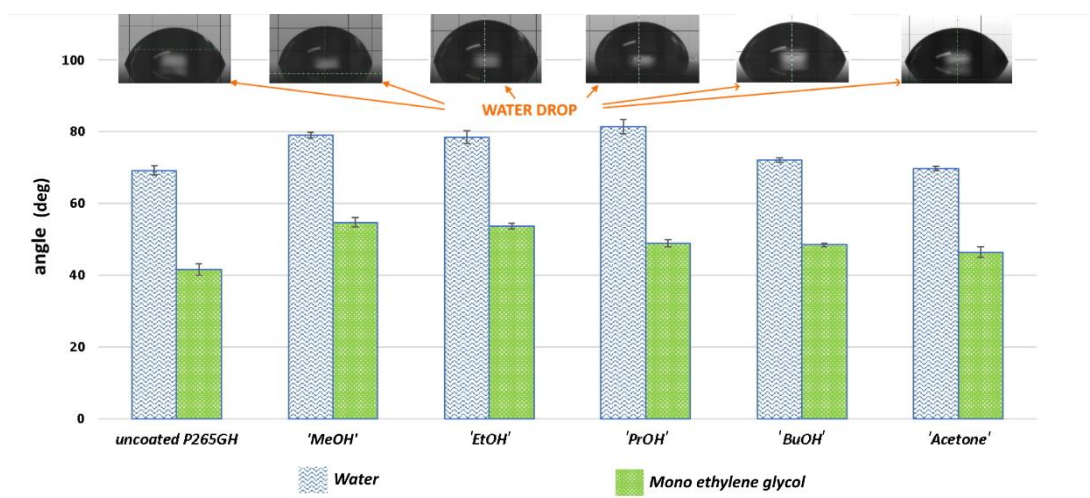

**Figure 8.** Contact angle values of water and mono ethylene glycol and images of water drops on uncoated P265GH steel and sol-gel coatings based-on various solvents

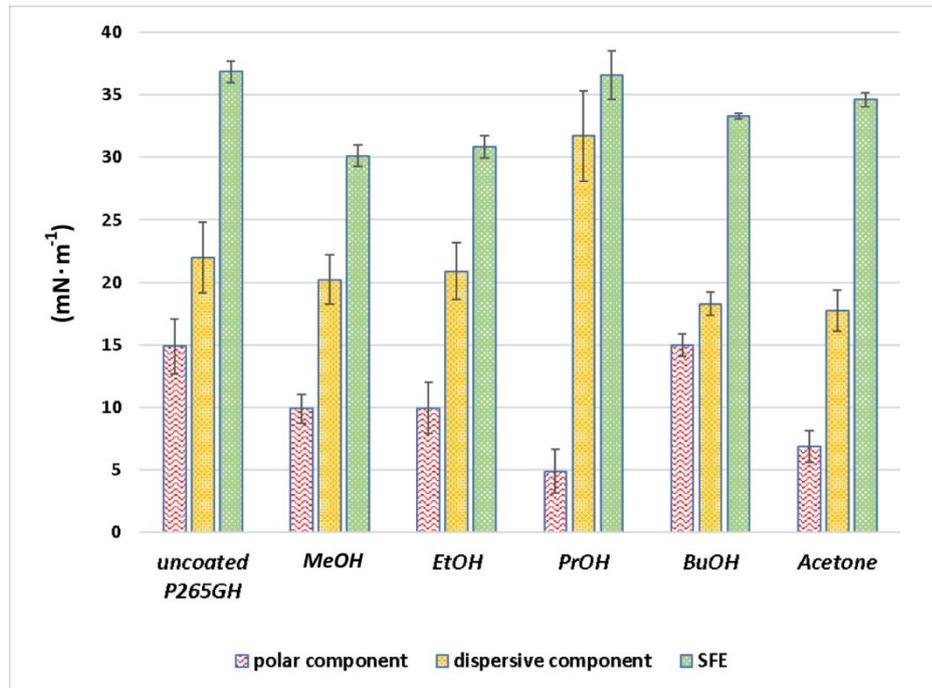

**Figure 9.** Surface free energy (SFE) with polar and dispersive components for uncoated P265GH steel and sol-gel coatings based on various solvents

According to literature data, friction stimulates delamination and increasing friction causes decrease in critical loads (LC) [49]. However, situation is not consistent for the samples presented. The results, for all presented coatings (Fig. 10), of the scratch test illustrated that the friction coefficient (approximately 0.5) and a friction force (approximately 8.5 N) are similar (Fig. 11), whereas the critical loads are significantly different between the obtained materials. For acetone-based coatings, the ($L_{C3}$) force is near 40% higher than that of methanol-based coatings (Fig. 10). Also, the size and extent/degree/amount of damage, especially observed for force $L_{C2}$, seems to differ from one type to another, from coatings based on ethanol and butanol, the tracks at the point of $L_{C2}$ are wider than other coatings and an increasing propagation of cracks along the tracks and chipping on the edges is observed. $L_{C2}$ forces, connected with cohesive failure (spallation), were observed in the range of 1.65-5.26 N. For all samples, the $L_{C1}$ force was in the range of 1.01- 1.79 N, for which circular Hertz type damage was observed. This damage developed along parallel orbits, with evidence of tensile cracking [49]. For coatings based on ethanol, the $L_{C1}$ force was additionally accompanied by evidence of chipping on the track edges. However, for coatings based-on propanol, the cracks had a different appearance, observed as a mild tensile crack.

Evidence of fragile cracking was not observed during the acoustic emission tests performed, , indicating that the observed damage has an elastic-plastic character. For the analyzed coatings, the damage observed in the coatings may initiate from decohesion between the coating and the P265GH steel substrate, observed as spallation at the first step of the test, and then followed by adhesion damage, such as is observed by the delamination of the coatings in the middle of the scratches. These results, along with the fact that all surfaces before coating were treated in the same way, suggest that, in the cases analyzed, changes in the structure of the coating materials and the elastic-plastic stress related to these have a significant influence on critical forces operating within the coating layers [48,50]. It should also be noted that the friction coefficient slowly increases during the increase in the normal force applied during the scratch test. This may be as a result of particles released from the chipping and cracking of the coating sticking to the scratch indenter and interfering with its function/performance, as presented by Blees et al. [51].

Changes within oxide matrix structure can have a significant influence on the scratch resistance of the coatings, as mentioned above. The sol-gel coatings based on siloxanes bond to the metallic substrate by interfacial bonding, based on hydrogen bonding, condensation reactions, and covalent bonding with hydroxyl groups localized on the surface of the steel substrate [48,50,51]. Furthermore, the mechanism of bonding of sol-gel materials to the metallic surface is usually explained by the conversion of hydrogen bonding via covalent condensation, which is promoted by thermal treatment of the coatings after their application [20].

According to Marcoen et al. [48], amine moieties from the silica precursors can promote interfacial adhesion by creating Lewis-type and Bronsted-type acid – base interactions between the free electron-pair of nitrogen (from the amine groups within the coatings) and iron (from steel). This fact can explain the highest value of $L_{C1}$ obtained for the coating based-on acetone as a solvent, for which the degree of conversion of amine groups from ApTEOS with the epoxy groupings from GPTMS) was the lowest (illustrated by the band at 648 cm$^{-1}$ in Fig. 5). Coating materials that consist of a substantial content of amine groups that are not bonded with epoxy ring of the GPTMS moiety can create a greater level of interaction with the steel surface. For the coating based-on acetone, all critical forces are highest measured of all of the coatings tested, and it appears that the presence of unreacted amine groups influence not only the adhesive properties of the bonding but also the cohesive properties as well. Additionally, it is believed that temperature treatment above 100°C promotes the formation of covalent bonds over the formation of hydrogen bonds [48], so covalent bonding mainly determines the scratch resistance properties of the obtained coatings.

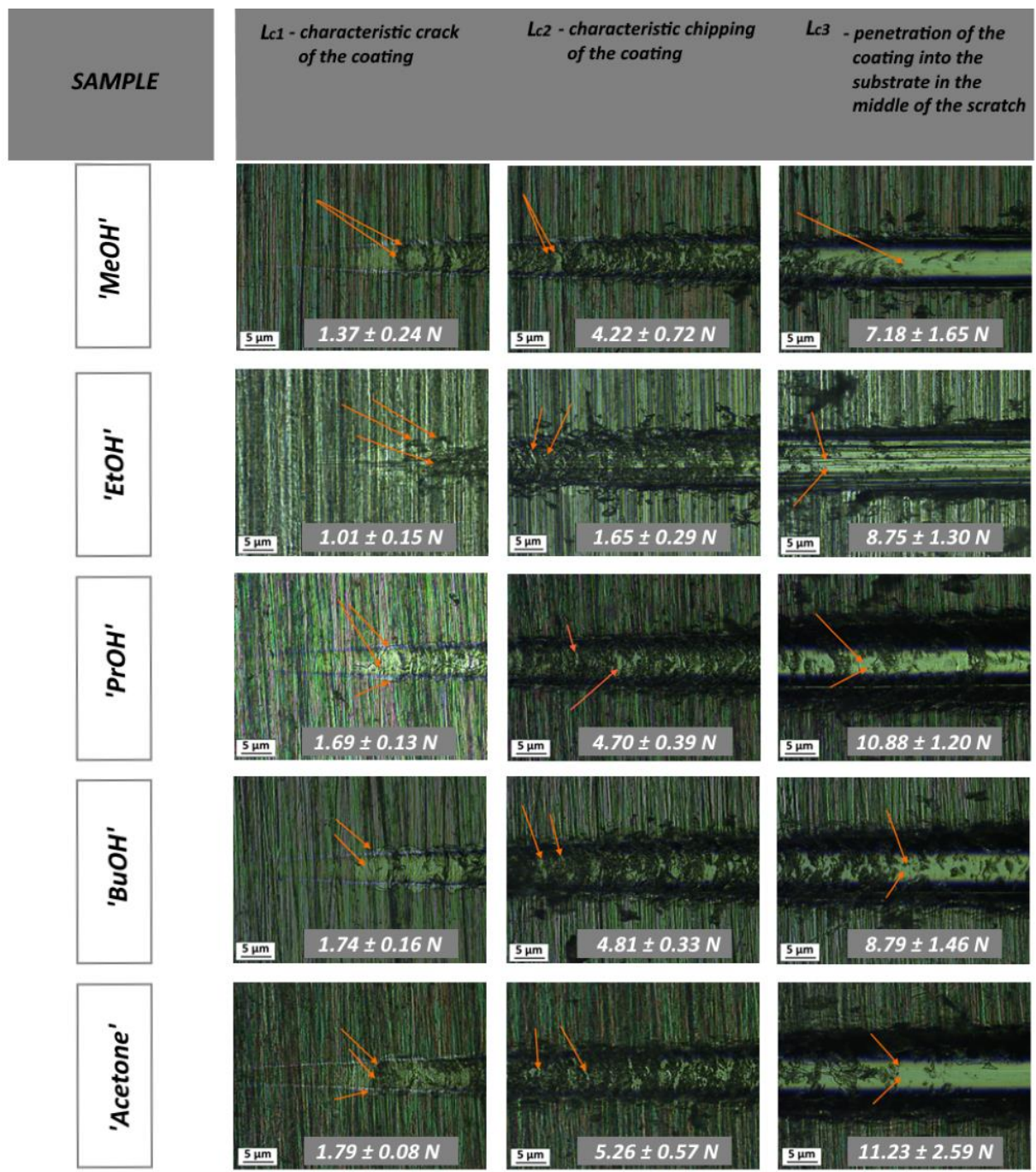

**Figure 10.** The results of the Scratch Test with characteristic loads

Fig. 11 shows: a) the trend of tangential force vs. normal force, where the tangential force increases linearly, thus increasing the progressive normal force for all coatings. For coatings based on organic solvents with longer hydrocarbon chains, such as propanol and butanol, and aprotic acetone, a decrease in tangential force is observed, which suggests that coatings based on these solvents are characterized by better lubricant properties that favor the sliding between the tip and the coating materials [50]. This seems to be confirmed by the evolution of the friction coefficient during the test (Fig. 11 b). For coatings based on propanol, butanol, and acetone, similar changes were observed for all scratches during the scratch test. The highest changes were observed for coatings based on using ethanol as a solvent. This result seems to correlate with critical loads, especially with $L_{c2}$ and the chipping influence. For the ethanol-based solvent coating, the chipping of coatings occurred relatively quickly (around $1.65 \pm 0.29$ N). Fig. 11 b) shows a relatively linear increase in the friction coefficient with increasing test load. Furthermore, the maximum friction coefficients for the analyzed coatings were around 0.4 at the end of the scratches.

A relatively low value of friction coefficient can be indicative of the presence of a significant organic component in the coatings [52], especially polymerized epoxide containing amine moieties. Fig. 11 c) and d) present the data for penetration depth ($P_d$) and residual depth ($R_d$) versus normal force. The largest $P_d$ and $R_d$ were obtained for the ethanol-based coatings, despite the greatest thickness, suggesting that using ethanol as a solvent, results in more efficient condensation, consuming the -OH groups created during hydrolysis. The interaction between -OH groups from the oxide network and the surface (Fe-OH) of the steel substrate generates an effective bonding at the interface between the coated material and the steel surface [52]. However, impoverishment in the level of -OH groups in the coating materials can result in decreased adhesion to the metallic substrate.

The plastic factor ($P_f$) for the coatings is shown in Fig. 11 e). The $P_f$ value describes the degree of plastic deformation of the coatings, and is calculated from the ratio of $R_d$ to $P_d$, according to equations 2-3 [50]:

$$P_f = 1 - E_f \qquad (2)$$

$$E_f = (P_d - R_d)/P_d \qquad (3)$$

where:

$E_f$ – plastic recovery (Fig. 11 f)

Due to the high value of the observed permanent deformation of the residual scratch track around 13 μm, and the value of the plastic factor around 1.0, it is concluded that the coatings are characterized by plastic deformation that is accompanied by very low elastic recovery [50].

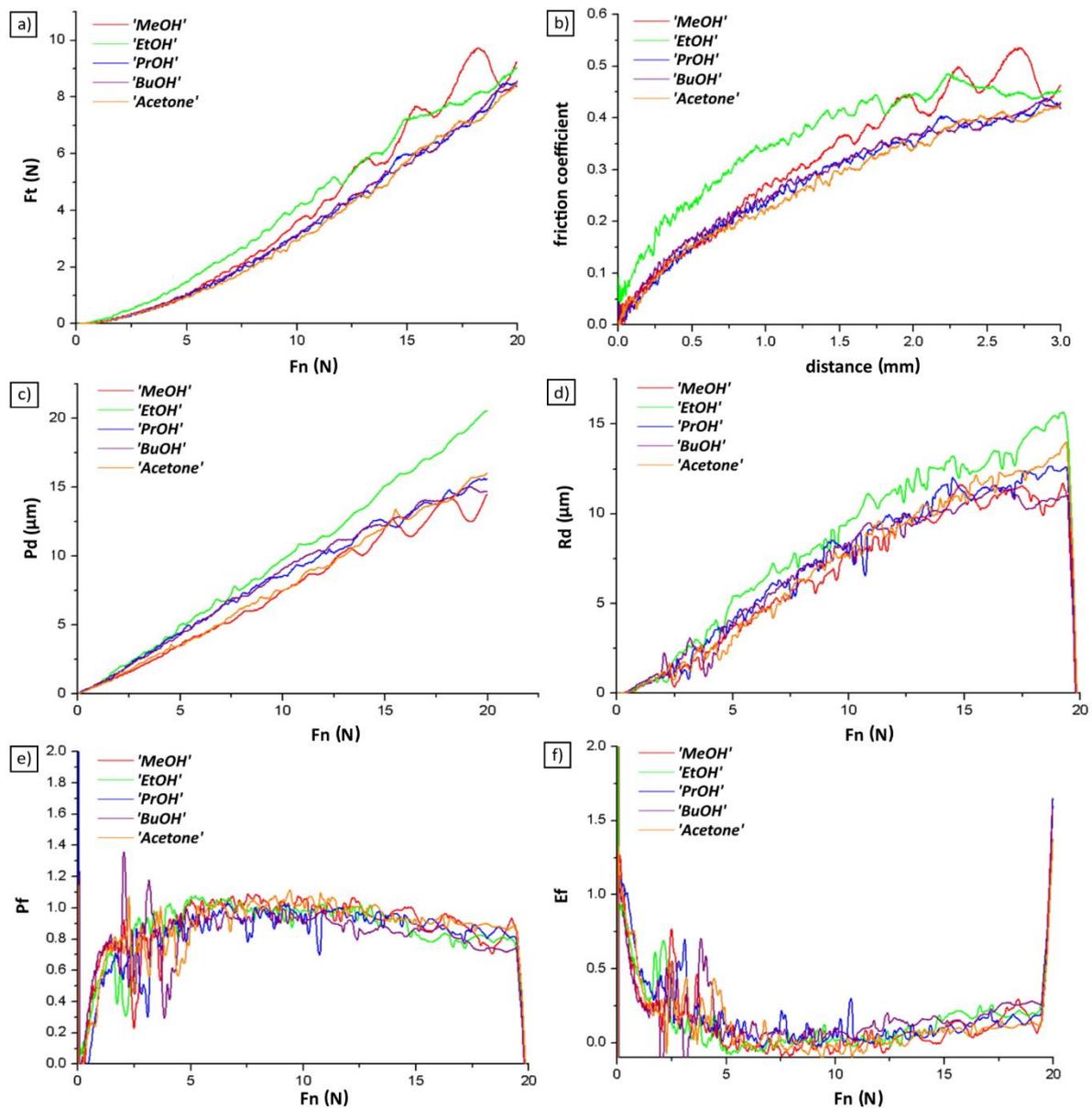

**Figure 11.** A) Tangential force evolution vs. normal force, b) friction coefficient evolution vs. scratched distance, c) trends of penetration depth (Pd), d) trend of residual depth (Rd), e) trend of plastic factor, and f) trend of elastic recovery for all coatings based on the various solvents used to prepare the surface coatings in the progressive load Scratch Testing mode.

Fig. 12 shows: a) the open circuit potential (OCP) change in time for uncoated P265GH steel and sol-gel coatings based-on various solvents in 0.5 M NaCl solution. Fig. 12.b) shows Tafel plots for the analyzed samples, respectively.

All analyzed samples were characterized by a decrease in $E_{OCP}$ during the initial exposure to 0.5 M NaCl. However, the plots for all of the sol-gel coatings move to more positive potential values than the uncoated P265GH steel, suggesting that coated surfaces are characterized by higher corrosion resistance relative to the control uncoated P265GH steel in 0.5M NaCl [53].

For Tafel plots, the anodic Tafel slope ($B_a$) and cathodic Tafel slope (Bc), the corrosion current densities ($J_{corr}$) and the polarization resistance ($R_p$) were determined and are presented in Table 4. The corrosion protection efficiency (η%) was calculated using $J_{corr}$ - equation (4) [54]:

$$\eta\% = \frac{Jcorr\ 0 - Jcorr\ p}{Jcorr\ 0} \times 100\% \qquad (4)$$

where $J_{corr}^0$ represents the current densities of uncoated P265GH steel, and $J_{corr}$ represents the current densities of the sol-gel coatings.

Based on the results obtained, higher values of $R_p$ were obtained for all surface-coatings, relative to the uncoated P265GH steel, although the increase for the coating based-on methanol was two orders of magnitude less than for the coatings prepared with the other solvents. Appropriately, coating by sol-gel materials caused a decrease of almost one order of magnitude in the $J_{corr}$ value, with respect to the uncoated steel substrate. The protection efficiency (η %) of almost all coatings was higher than 90%, suggesting that these coatings can effectively protect P265GH from corrosion in the environment of a 0.5 M sodium chloride solution [7,54]. However, the lowest $R_p$, $J_{corr}$ and η% values were observed for the butanol-based solvent coatings. These results could be a consequence of defects and the cracks observed in the coatings via SEM. These defects prevent the coating from performing as a successful barrier against corrosion agents such as Cl$^-$, since they expose the substrate to the corrosive agent. For uncoated P265GH, a higher value was obtained for the anodic Tafel slope ($B_a$) compared to the cathodic slope (Bc), indicative of Fe dissolution of Fe. Dissolving Fe can additionally influence the $E_{OCP}$ [55]. A decrease in $B_a$ was observed for all of the sol-gel coatings, suggesting that the anodic reaction was inhibited. Furthermore, all Tafel curves (Fig. 12 b) showed an anodic shift, with displacements to more positive potential values, meaning that sol-gel coatings mainly prevent anodic dissolution. A similar conclusion for coatings based on GPTMS has been reported by Balaji et al. [54].

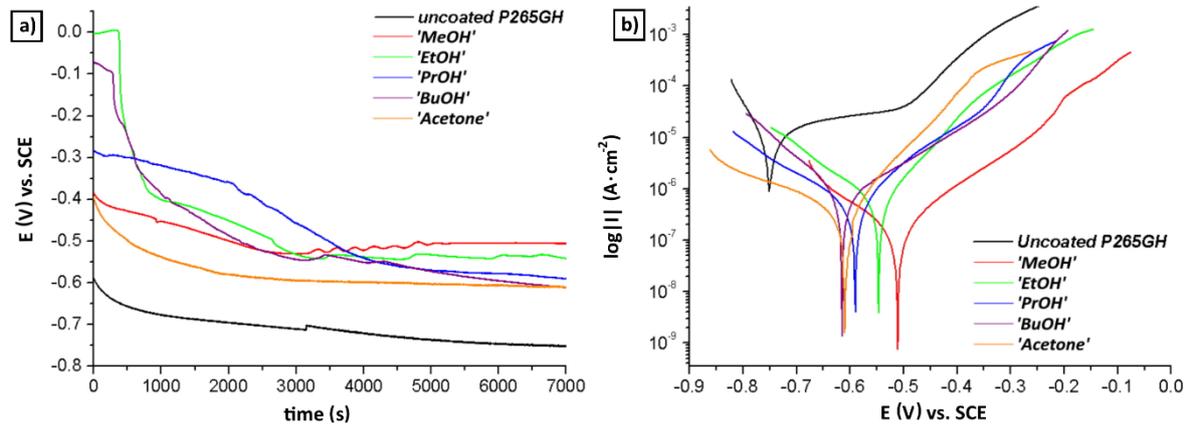

**Figure 12.** The OCP curves a) and Tafel plots b) for sol-gel coatings based-on various solvents

**Table 4.** Corrosion performance parameters for uncoated P265Gh steel and sol-gel coatings

| SAMPLE | $B_a$ (mV· dec$^{-1}$) | $B_c$ (mV· dec$^{-1}$) | $R_p$ (k$\Omega$· cm$^2$) | $J_{corr}$ ($\mu$A· cm$^{-1}$) | η % |
|---|---|---|---|---|---|
| *Uncoated P265GH* | 544.8 ± 110.2 | 97.9 ± 20.6 | 3.83 ± 0.88 | 9.41 ± 1.29 | - |
| *'MeOH'* | 122.5 ± 32.4 | 203.9 ± 167.6 | 138.58 ± 27.40 | 0.23 ± 0.04 | 97.6 |
| *'EtOH'* | 95.4 ± 17.3 | 328.9 ± 167.6 | 44.50 ± 14.13 | 0.73 ± 0.22 | 92.2 |
| *'PrOH'* | 136.5 ± 5.4 | 287.0 ± 34.8 | 40.59 ± 12.04 | 1.03 ± 0.32 | 89.1 |
| *'BuOH'* | 266.6 ± 24.3 | 224.8 ± 63.1 | 24.89 ± 9.67 | 1.79 ± 0.09 | 81.0 |
| *'Acetone'* | 109.5 ± 3.5 | 433.6 ± 136.6 | 63.46 ± 21.02 | 0.70 ± 0.22 | 92.6 |

Electrochemical impedance spectroscopy (EIS) was used to assess the performance of the coatings against corrosion after exposure to various corrosive solutions. The tests were carried out after 2 h of immersion in 0.5 M NaCl medium, and the corresponding Nyquist and Bode plots are shown in Fig. 13. The data of the experimental results presented in Table 5 have been fitted using ZSimVin software, and the corresponding equivalent circuit proposed is presented in Fig. 13) The equivalent circuit consists of a solution resistance ($R_S$), a constant phase element connected with the sol-gel coating ($CPE_{SG}$), resistance of the sol-gel coating ($R_{ST}$), capacitance of the corrosion product ($C_{CP}$), resistance of the corrosion product ($R_{CP}$), a constant phase element connecting with the double electric layer ($CPE_{DL}$) and a charge transfer resistance ($R_{ST}$), and this circuit was used to determine EIS value parameters. Due to the complicated nature of the corrosion process as well as surface roughness and inhomogeneities, the capacitor within the in EIS circuit often does not behave ideally. The impedance of CPE is represented by a shift from an ideal capacitor, defined by equations (5-6) [55]:

$$Z_{CPE} = [Y_o(j\omega)^n]^{-1} \quad (5)$$

$$C = Y_o(2\pi f_{max})^{n-1} \quad (6)$$

where: $Y_0$-CPE is a constant, J represents an imaginary number, $\omega$ represents the angular frequency ($\omega=2\pi f$), and n is the exponent of CPE $-1<n<1$.

For n=-1, the CPE represents an inductor, when n=0 it is a resistor, and a value of n=-1 denotes a capacitor [56].

$f_{max}$ – is the frequency at which the imaginary value reaches a maximum on the Nyquist plot

Two parallel circuits with two time constants were used to fit the EIS curves obtained for the coatings based on methanol and acetone [52], These were connected with $R_{SG}$ and $CPE_{SG}$ at high frequency (HF), between 10-300 Hz, and $R_{CT}$, and to $CPE_{DL}$ at low frequency (LF), near 0.1 Hz. For both coatings in contact with water, $R_{SG}$ was relatively higher than the $R_{CT}$ response of the layer close to the interface between the coating and the steel. However, for coatings based on acetone, the relation between $R_{CT}$ to $R_{SG}$ increased, arising from superior bonding of the coatings to the steel substrate. These results correlate to the findings from the Raman spectra and the Scratch Test. For the sol based on acetone as a solvent, band at 1640 cm$^{-1}$ comes from NH$_2$ moiety from unreacted ApTEOS, was observed (Fig. 5). Unreacted NH$_2$ groups promote bonding with the P265GH steel surface [48], which can be observed by the obtained $L_{C3}$ value, which was also the highest for the acetone-based coating, compared with all the other coatings presented. On the other hand, for coatings based on methanol as a solvent, the $R_{SG}$ is nearly one order of magnitude greater than the others. The higher physical barrier against corrosion of methanol-based coatings can be due to the greater homogeneity and compactness of this coating, which was confirmed by the lowest $CPE_{SG}$ obtained, being connected with the lowest water absorption by the methanol-based coatings in comparison with the other sol-gel coatings [57], having also the highest $R_{SG}$ of all of the coatings. Additionally, an enhanced corrosion resistance of the methanol-based coating was manifested by the increase in the impedance modulus |Z| [57,58], which was one order of magnitude higher at LF than observed for the other coatings. The phase angle, LF, which describes the electrochemical activity occurring inside the defect formed at the interface [59] was highest for the methanol-based coating, reflecting the interface without defects and the high corrosion resistance of this coating. For the coatings based on ethanol, propanol and butanol as solvents, a two RC circuit model also applied. What is more, for the equivalent circuits used for these three coatings, the appearance of a parameter characteristic for the formation of corrosion products ($R_{CP}$ and $C_{CP}$) was observed

[60]. Also, the time constants (capacitive loops) for these coatings shifted to a medium frequency (MF), suggesting that, during the coating, corrosion products were created [58,60]. However, for the coating based on ethanol as a solvent, a more nested equivalent circuit was observed, suggesting that the corrosion products were more compact than for coatings based on propanol and butanol as synthesis solvents. This result deviates from the assumption that a thicker coating provides a better barrier against corrosion agents, as in other types of coatings [61,62], and suggests that ethanol-based coatings may contain defects within their volume. The relatively high phase-angle observed in the MF for these coatings that used ethanol, propanol, and butanol as synthesis solvents may be the result of the corrosion products formed at the interface between the coating and the metal [58,60]. However, the corrosion layers that were formed did not act as successful barriers and, during exposure to corrosion medium, the corrosion protection decreased [56,60].

The Nyquist plots (Fig. 13) for the coatings prepared with ethanol-, propanol- and butanol as synthesis solvents displayed a decrease in the impedance modulus (|Z|) by one order of magnitude, compared to the other coatings presented. Also, an increase in the $CPE_{SG}$ and decrease in the $R_{SG}$ were observed for the coatings prepared with these three solvents, and, in each case, this was followed directly by adsorption of the electrolyte and an associated loss in the protective properties of these coatings [52,63]. According to the circuits used in the EIS fitting circuits for propanol and butanol, two weak separated constants in LF were observed. However, the character of these elements was different than for the ethanol-based coatings. Here a parallel circuit, consisting of a separated circuit arising from the corrosion product, suggested that the corrosion film was formed as a loose form [58].

Based on the SEM results for coatings based on propanol and butanol, where defects and cracks were observed, the coatings showed channels that could serve to migrate corrosion attack of the underlying steel substrate by agents and explain the inferior corrosion protection properties of these coatings. Corrosion attack of the metallic substrate coated with sol-gel coatings based on propanol and butanol as synthesis solvents, were not effectively inhibited, since these coatings were characterized by defects, which provide access of the corrosion agents to the metallic surface.

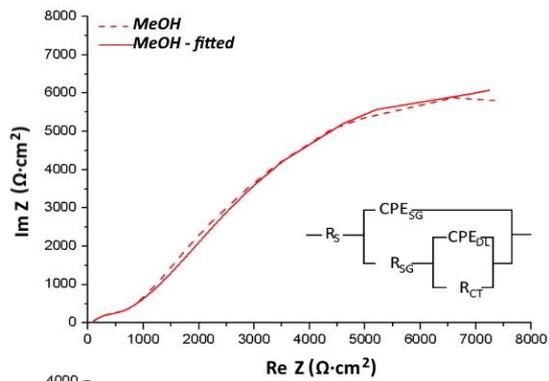
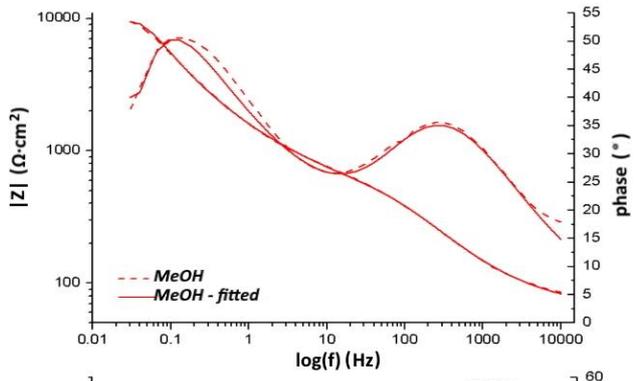
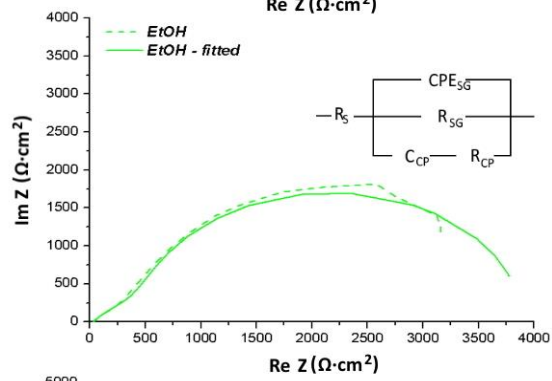
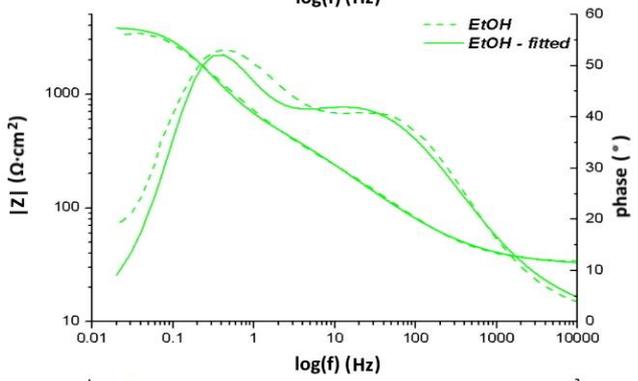
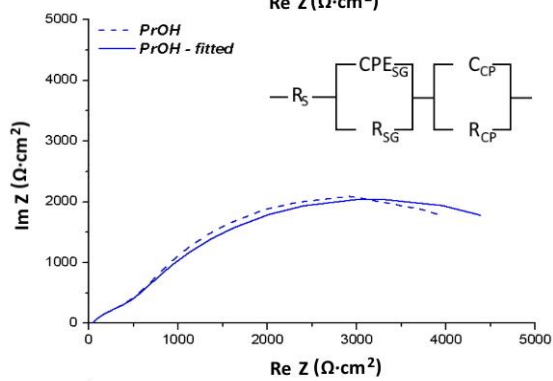
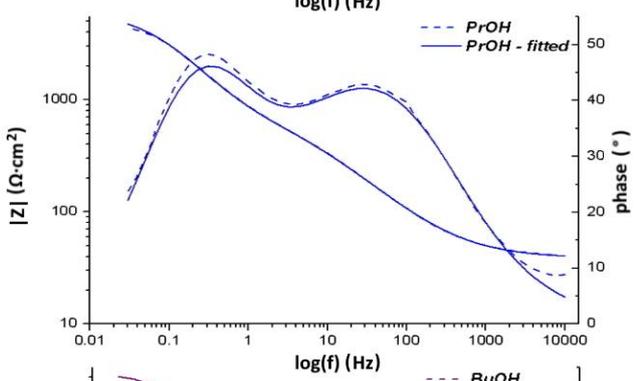
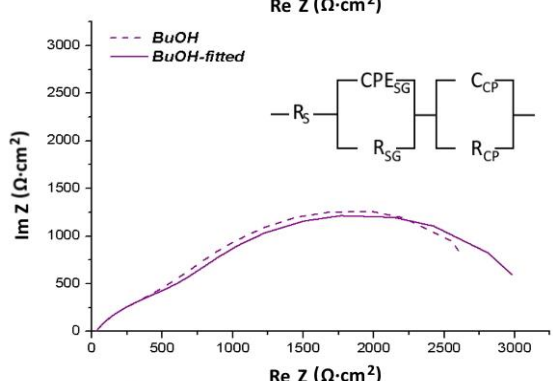
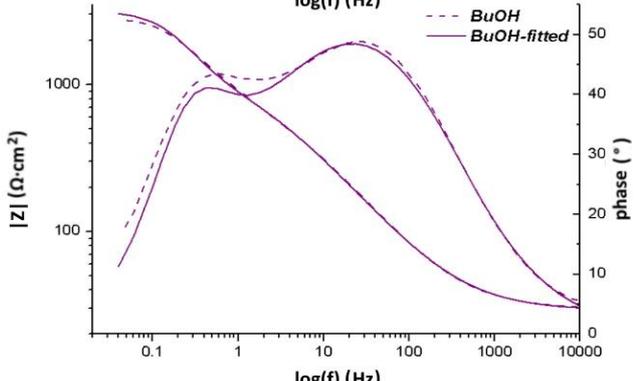
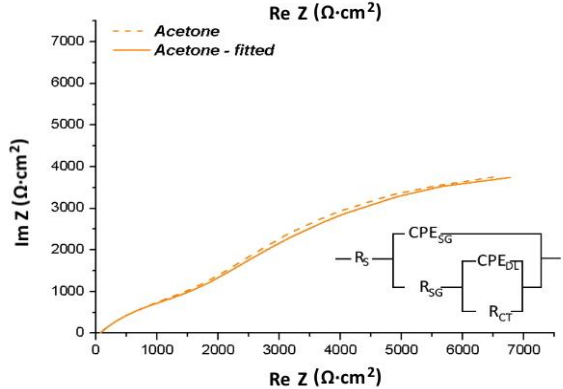
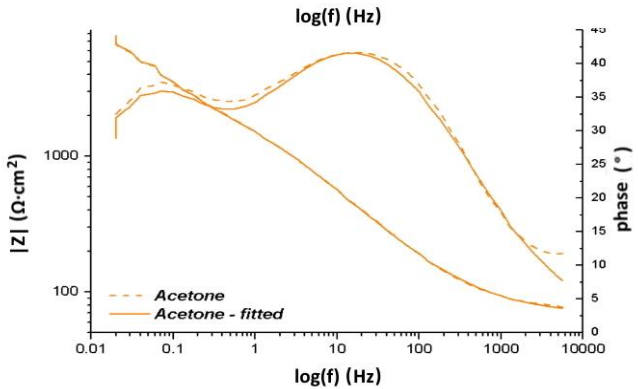

**Figure 13.** EIS study for the obtained coatings: Nyquist (left) and Body (right) plots.

**Table 5.** EIS performance parameters for sol-gel coatings based on various solvents

| SAMPLE | $R_s$ ($\Omega \cdot cm^2$) | $CPE_{SG}$ ($\mu S \cdot cm^{-2}$) | n | $R_{SG}$ ($k\Omega \cdot cm^2$) | $CPE_{DL}$ ($\mu S \cdot cm^{-2}$) | $C_{CP}$ ($mF \cdot cm^{-2}$) | n | $R_{CT}$ ($k\Omega \cdot cm^2$) | $R_{CP}$ ($k\Omega \cdot cm^2$) |
|---|---|---|---|---|---|---|---|---|---|
| *'MeOH'* | 65.2 ± 4.7 | 4.29 ± 0.9 | 0.62 ± 0.01 | 31.95 ± 0.96 | 20.86 ± 4.76 | | 0.70 ± 0.02 | 0.898 ± 0.098 | |
| *'EtOH'* | 36.0 ± 2.6 | 15.9 ± 3.0 | 0.64 ± 0.02 | 3.739 ± 1.386 | | 17.3 ± 0.03 | | | 1.197 ± 0.206 |
| *'PrOH'* | 33.5 ± 6.8 | 18.8 ± 5.0 | 0.68 ± 0.02 | 3.736 ± 2.279 | | 44.5 ± 5.5 | | | 0.710 ± 0.277 |
| *'BuOH'* | 30.0 ± 1.4 | 22.3 ± 5.9 | 0.67 ± 0.04 | 2.198 ± 0.912 | | 44.7 ± 3.2 | | | 0.734 ± 0.193 |
| *'Acetone'* | 70.0 ± 30.8 | 11.4 ± 5.7 | 0.61 ± 0.02 | 10.176 ± 1.821 | 30.6 ± 8.0 | | 0.65 ± 0.08 | 2.229 ± 0.477 | |

*Discussion*

The sol-gel syntheses described in this study used five different solvents. At this point, the influence of solvents on the structure sol-gel materials should be discussed. Methanol, ethanol, propanol, and butanol are classified as protic solvents, while acetone is aprotic [24] – characteristic properties are presented in Table 2. On the other hand, the selection of precursors, in this case GPTMS and ApTEOS, and the by-products of the hydrolysis and condensation reactions, such as methanol, ethanol, and water, can play a key role in determining the overall direction of the synthesis. An aprotic solvent enables the creation of hydrogen bonding with anions, since free electron-pairs attract cations, which positive promotes hydrolysis. Mechanism based on substituting the $C_2H_5^+$ released from the ApTEOS moiety and the $CH_3^+$ released from GPTMS by $H^+$ promote the creation of Si-OH [48]. Additionally, in the case of methanol and ethanol solvents (these alcohols are also by-products from the hydrolysis and condensation processes), shifting the reaction equilibrium towards hydrolysis, according to Le Chaterier (Fig. 14), influences the kinetics of sol-gel process.

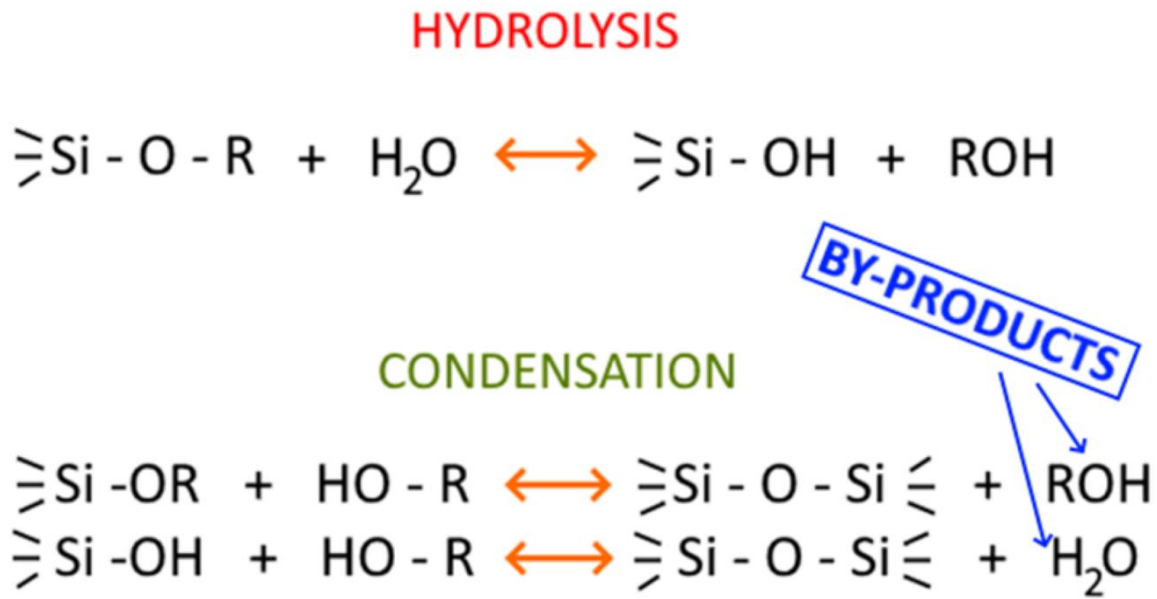

**Figure 14.** General hydrolysis and condensation reactions in sol-gel synthesis

A higher degree of condensation fosters the formation of a dense network with excellent protection properties. On the other hand, the polarity and dielectric constant of the solvent used can play a key role in driving sol-gel reactions to achieve specific types of networks. For the dissolved precursor substances and solvent, there exist two energetic barriers[14]. Solvents characterized by lower dielectric constants promote the formation of a dense network with greater structural stability and mechanical endurance. In turn, polarity is related to dipole-dipole interactions, which significantly influences the miscibility of the reaction components. In this, the miscibility of the used solvents with water (one of the by-products of the condensation reaction), which controls the condensation reactions, should be considered. Immiscible compounds, or those characterized by limited miscibility (e.g. butanol), can influence the change in condensation kinetics and, as a result, influence the nature and quality of the oxide network [14]. This fact can explain the worse protection properties observed with the coating based on butanol, despite the low dielectric constant of butanol, even though a low dielectric constant would suggest that a coating would be characterized by improved protection parameters. Also, it should be noted that, during coating stabilization, greater solvent molecules promote formatting a greater size porous [20], which are channels for the corrosion agent to migrate to the substrate surface. This could explain why coatings based on ethanol as a solvent are characterized by protection properties that are worse than those based on methanol, despite the greater thickness of the coatings and, it would seem, a high affinity for by-products. On the other hand, the quantity and quality of the functionalizing moieties (e.g. epoxy ring, amine group), influence on rate of hydrolysis. Hydrophobic silane hydrolyzes at a slower rate [47]. Furthermore, the stronger basicity of the silane precursors substitutions significantly accelerate the kinetics of hydrolysis [64]. In this study, the methoxide substitutions from GPTMS are stronger than the ethoxide moieties from ApTEOS. In addition, the smaller hydrolyzing groups increase the kinetics of hydrolysis. As a result, the speed of hydrolysis is correlated with the steric effect of the alkoxyl groups. With the above in mind, and the fact that among the precursors used for sol-gel reaction, GPTMS has smaller hydrolyzing moieties, namely, methoxy moieties, it is expected that GPTMS could hydrolyze faster than ApTEOS. It expectedthat the quantity of Si-OH groups, which are accountable for the hydrophilic effect and increase the coating adhesion to the metallic substrate, should be controlled by ApTEOS hydrolyzing. However, bearing in mind that the variable factor in the current study was the solvent used, seems to solvation effects, and so steric hindrance of solvent, can control hydrolysis kinetic materials consisting of GPTMS and ApTEOS. The prototic solvent with a relatively lower CH unit ratio should promote the hydrolyzing of siloxanes groups, which can suggest an increase in the degree of cross-linking of sols based on methanol and ethanol and can be observed by decreasing the SFE of the modified surface and increasing $R_p/R_{SG}$. On the other hand, the question about interfacial bonding availability and interaction between unhydrolyzed moieties like epoxy groups from GPTMS and amine from ApTEOS, still remain. Amine groups cause the opening of the epoxy ring in GPTMS [37,38,48], contributing to the increase in the densities of the materials. However, amine groups can also form bonds with the metallic substrate to which the coating is applied. Despite the fact that the covalent bond -M-OSi created between steel and the sol-gel oxide coating is stronger than amine-steel, additional moieties capable of creating bonds with the metallic surface can increase adhesion. It can be observed via example of the acetone-based coating, where a characteristic band was observed at 648 cm$^{-1}$ for the sol form was observed with Raman spectroscopy, indicative of a free amine group that remains unreacted with the epoxy ring and which can create a bond with the metallic surface. As a result, the highest value of $L_{c3}$ (critical load) was observed for this coating.

## 4. Conclusions

The use of GPTMS and ApTEOS, as precursors in sol-gel reactions makes it possible to obtain coatings with properties that are strongly dependent on the nature of the solvent used in the synthesis. Solvent parameters, such as polarity, dielectric constant, miscibility with water, and the by-products of hydrolysis and condensation, play a significant role in the end-point synthesis of sol-gel materials. Coatings based on methanol are characterized by near 98% corrosion protection efficiency and the highest resistance associated with sol-gel coating (nearly one order of magnitude greater than observed with the other solvents presented). The greatest coating thicknesses were obtained with ethanol-based coatings. However, increased thickness did not correlate with increased corrosion protection. Coatings based on propanol and butanol solvents were characterized by cracks and, especially for coatings based on propanol, increased thickness of the corrosion films produced via sol-gel coating. Acetone, as a reaction medium, influenced the interaction between amine groups of the ApTEOS and epoxy rings of the GPTMS, thereby increasing the availability of amino groups that increases the strength of bonding with metallic surface, thereby improving the adhesion of the coating layer to its substrate.

**Funding**

The research was partially supported by the National Science Centre, Poland under the OPUS + LAP project 'Research on the influence of self-healing, organic-inorganic sol-gel layers on the corrosion resistance and fatigue of steel in the VHCF range' UMO-2020/39/I/ST5/03493

Paper has been published in

**Ceramics International, Volume 48, Issue 24, 15 December 2022, Pages 37150-37163**

https://doi.org/10.1016/j.ceramint.2022.08.291

please cite this article as:

Jolanta Gąsiorek, Anna Gąsiorek, Bartosz Babiarczuk, Walis Jones, Wojciech Simka, Jerzy Detyna, Jerzy Kaleta, Justyna Krzak, Anticorrosion properties of silica-based sol-gel coatings on steel – The influence of hydrolysis and condensation conditions, Ceramics International, Volume 48, Issue 24, 2022, Pages 37150-37163, ISSN 0272-8842, https://doi.org/10.1016/j.ceramint.2022.08.291.